\documentstyle[12pt,psfig]{article} 
\font\elevenbf=cmbx10 scaled\magstep 1                                        
\begin{document}                                                              
%\begin{flushright}{LNF-94/xxx P}  
%\end{flushright}                                                      
\begin{center}                                                                
{ \large\bf Time delayed $K^+N$ reactions and exotic baryon resonances\\}
\vskip 2cm  
{ N. G. Kelkar$^1$, M. Nowakowski$^1$ and K. P. Khemchandani$^2$\\} 
$^1$Departamento de Fisica, Universidad de los Andes,
Cra.1 No.18A-10, Santafe de Bogota, Colombia\\
$^2$Nuclear Physics Division, Bhabha Atomic Research Centre, Mumbai 400085
\end{center}
\vskip .5cm
\begin{center}
\end{center}                             
\begin{abstract}
Evidences and hints, both from the theoretical and experimental side, of exotic
baryon resonances with $B=S$, have been with us for the last thirty years.
The poor status of the general acceptance of these $Z^*$ resonances 
is partly due to the prejudice
against penta-quark baryons and partly due to the opinion that a proof of the
existence of exotic states must be rigorous. This can refer to the
quality and amount of data gathered, and also to the analytical methods
applied in the study of these resonances. It seems then mandatory that all 
possibilities and aspects be exploited. We do that by analyzing the time delay
in $K^+N$ scattering, encountering clear signals
of the exotic $Z^*$ resonances close to the pole values found in partial wave 
analyses. 
\end{abstract}                                                                
\newpage 
\section{Introduction}
With the advent of the $N^*$ program at Jefferson Lab \cite{burkert} and the 
forthcoming Japan Hadron Facility \cite{japan}, the interest in baryon
resonances has increased over the last few years. Mysteries of veteran 
resonances, the hope to find new and `missing' ones and the study of model 
predictions like the Unitarized Chiral Perturbation Theory \cite{ucpt} are 
only some topics worth 
mentioning in this context. But of course, one of the most interesting and 
exciting subject is the quest to find exotic hadrons \cite{exotics}. 
Not all such hadronic states which we would call exotics,  
can be identified unambiguously. 
In the mesonic sector this would apply for
glueballs, hybrids and mesonic molecules with conventional
quantum number assignments. A sure candidate for an exotic
meson would be the so-called $J^{PC}$ exotic, 
which cannot be made up of quark anti-quark pair and
appears with quantum numbers, $J^{PC}= 0^{--}, 0^{+-}, 1^{-+}, 2^{+-}$ etc. 
Some candidates
have indeed been found \cite{JPCmesons}, but as in the case of exotic baryons
they are not yet fully accepted. 
In the baryonic sector, a clear example of an exotic baryon would be
one with its baryon number equal to the strangeness quantum
number, i.e $B=S$.
Indeed, such a hadronic state
would either have to be a composite of five valence quarks, i.e., 
$q^4\bar{s}$ with $ q\neq s$, or a $K^+ N$ molecule i.e. a bound
state of hadrons. These resonances are usually 
called $Z$ or $Z^*$ resonances. Evidence
for the existence of such resonances dates back to the early 70' s 
(see \cite{yogi, kato} for a collection of references on early $KN$ 
experiments) and 
continues in the 80's ( see the reference list in \cite{arndt, arndt2}). 
This has been accompanied   
by theoretical predictions mostly in favour of the existence 
of penta-quarks \cite{theory,roiesnol}.
The situation is, however, more than confusing as the experimental signals
are called `pseudoresonances', `doorways', `resonance-like structures' and
`resonance-like loops' (moving counterclockwise in the Argand diagrams). 
These terms do not carry much physical meaning. We say this because,  
a resonance is at least theoretically, clearly defined as 
an unstable state characterized by different quantum numbers.  
The physical reasons for this caution are also not
always clearly stated in the papers. However, in \cite{nakajima} we find the
following statement: ``Martin and Oades [2] [our reference \cite{martin}] 
interpreted these waves as complicated structures in the unphysical sheet
instead of simple resonance poles, because the peaks of the speed plots did 
not coincide with the peaks of the ``resonance''. This is also the case for 
the three waves obtained in the present paper.'' Since we have already done 
an analysis of time delay (a concept related to speed plots as
we shall explain below) \cite{ng, ngandme, ngandmeagain}, the above
statement has motivated us to look into the matter more closely.
Using the latest $K^+N$ scattering data in the form of 
phase shifts and $T$-matrix solutions \cite{arndt, single}, we calculate both 
the time delay and speed plots. 
We adopt the conservative point of view that we can promote eventually
a resonance candidate to a fully accepted member of the resonance spectrum,  
only if the pole values of the masses
found in the partial wave analysis agree with the values of peaks in 
speed plots and time delay. This restriction is more than one can impose 
on standard resonances. For example, the $S_{11}$ resonance $N(1535)$ 
which is claimed to be missing in the speed plots
\cite{hoehler} and the $P_{31}$ resonance $\Delta (1910)$ which gives no 
signal in the time delay plots \cite{ng, ngandme} are both taken as 
well-established resonances. Hence, our requirement for the confirmation
of an exotic $Z^*$ resonance is doubly strict.       
\setcounter{equation}{0}
\section{Time delay and speed plots}
In this section, we shall discuss the concepts of time delay 
and speed plots. Though both the methods are
useful tools in analyzing resonances, they are different
quantities (in certain cases, they differ only by a crucial 
minus sign) and their origin is also different. Time delay has an
intuitive background which we would like to explain. The authors of
\cite{bjorken} in a section where they compare Feynman diagrams to
electric circuits state: ``If it is possible for the intermediate 
particle to be {\it real}, then the process becomes unbounded in space-time and
the corresponding amplitude singular''. If this intermediate state 
(formed for example in two body scattering) is in the 
$s$-channel and is also a resonance, the singular amplitude
can be tamed by a Breit-Wigner form, at least for elementary processes like
for example the $e^+ \, e^- \,\rightarrow Z^0\rightarrow\, \mu^+ \, \mu^-$ reaction.  
A similar `catastrophe' can take place with a particle in the 
$t$-channel if one starts with an unstable particle \cite{tchannel}, 
but this happens because we have violated the
requirement that the scattering states be asymptotically free at large 
distances from the scattering centre. In any case, both reactions are 
non-localized (`unbounded') in space-time. The first case which is of interest
for us here, can be visualized as follows: a resonance is produced on-shell
at a space-time point $(t_1, x_1)$, it propagates for some time and decays
at a space-time point $(t_2, x_2)$. Certainly, the reaction is time delayed
(by an amount $\Delta t = t_2 - t_1$)
and the time delay has to be {\it positive}.  This picture is 
qualitatively model independent, as it does not depend 
on the form of the analytical tool by which the resonance is
described (Breit-Wigner, a modified version of the same, etc.) and this 
makes it useful for broad hadronic states.  
Eisenbud and Wigner \cite{wigner, wigner2, wigner3} found a way to 
evaluate this time delay, $\Delta t$, from the phase shift $\delta$ as, 
\begin{equation}\label{1}
\Delta t = 2 \hbar {d\delta \over dE}\,.
\end{equation}
It can also be evaluated from the $S$ matrix, generalizing at the same time, 
the time delay in elastic channels (\ref{1}) to an arbitrary 
reaction $i\to j$ as, 
\cite{smith}
\begin{equation}\label{2}
\Delta t_{ij} = \Re e \biggl [ -i \hbar (S_{ij})^{-1} {dS_{ij} \over dE}
\biggr ] \, 
\end{equation}
with the identification, $\Delta t_{ii}= \Delta t$. Defining the $T$ matrix as
\begin{equation}\label{3}
S_{kj} = \delta_{kj} + 2 i T_{kj} \,,
\end{equation}
with
\begin{equation} \label{3a}
T_{kj} = \Re e T_{kj} + i\Im m T_{kj}
\end{equation}
we get  
\begin{equation}\label{4}
S^*_{ii} S_{ii}\, \Delta t_{ii} = 2 \hbar \biggl[ {d\Re e T_{ii} \over dE}
+ 2 \Re e T_{ii}
{d\Im m T_{ii} \over dE} - 2 \Im m T_{ii} {d\Re e T_{ii} \over dE} \biggr].
\end{equation}
Parameterizing a general $T$-matrix for the elastic channel in the presence
of non-zero inelasticities, through the phase shift $\delta$ and
the energy dependent inelasticity parameter $\eta$ as, 
\begin{equation} \label{5}
T={\eta e^{2i\delta} -1 \over 2i},\,\,\,\,\,\, 0 < \eta \leq 1
\end{equation}
one can show that $\Delta t$ is given by (\ref{1}) even if $\eta \neq 1$.
Time delay can be positive as well as negative. A big positive peak is expected
within the kinematical vicinity of a resonance. Large regions of negative
time delay can occur due to the 
opening of new channels or due to a repulsive interaction, or in the 
presence of several resonances even with all inelasticities zero 
\cite{lipschutz}.
Sometimes one hears/reads statements like `the phase shift has to change
sharply to indicate a resonance' or `a phase motion indicates a resonance'.
Very often it is stated that 
the phase shift has  to increase by an odd multiple of $\pi/2$
while passing through the resonance region. Whereas the first statement is not
precise enough, the second one is model dependent. Indeed, it originates from
\begin{equation} \label{6}
\delta_{res}(E) \simeq \tan^{-1}\biggl [{\Gamma /2 \over E_R -E} \biggr ]
\end{equation}
which gives a Breit-Wigner form, namely, 
\begin{equation} \label{7}
{d\delta \over dE} = {\Gamma \over 2} {1 \over (E_R-E)^2 +\Gamma^2/4} \,.
\end{equation}
The reason behind both the above statements is actually the time delay, 
which is a model independent analytical justification of both of them. 
Moreover, as evident from the last equation, $\Delta t(E)$ is 
essentially the spectral density used to calculate survival 
probabilities \cite{spectral} which
carries some importance if $\Delta t$ is not a Breit-Wigner.

The concept of time delay and its connection to resonances is well documented
in many papers and found its entry in many textbooks. For a complete list
of references we refer the reader to \cite{ngandmeagain}. Here we remark that
an operator for time delay has been found by Lippmann in \cite{Lippmann}. 
Time delay can also be applied to steady-state solutions of 
Maxwell equations for the total reflection case \cite{agudin}, 
to chaotic scattering
\cite{chaos} and to transport theory in heavy ion collision \cite{qgp}, 
all showing the wide applicability of $\Delta t$. Before
applying it to $KN$ scattering, we would like to compare it with speed plots.  

The speed plot is defined through
\begin{equation} \label{8}
SP(E)= \biggl | {dT \over dE}\biggr |\,.
\end{equation}
Its first appearance is less clear than in the case of time delay. It probably
stems from the Argand diagrams, as it describes the speed at which the curve
in the Argand diagram is traversed
with $E$ playing the role of the affine parameter. 
Using equation (\ref{5}) we get, 
\begin{equation} \label{9}
SP(E)=\sqrt{\eta^2\biggl ({d\delta \over dE}\biggr )^2+{1 \over 4}
\biggl ({d\eta \over dE}\biggr )^2}
\end{equation}
which shows that if $\eta = 1$, $SP(E) = \vert \Delta t \vert /2$.
Even in this case, a speed plot is not the same as a time delay plot, 
since large negative regions 
in the time delay plots will become positive peaks in the speed plots.
This fact is unfortunate, since only positive peaks in time delay 
indicate a resonance.
For example, the negative dips in time delay in the $S_{01}$, 
$D_{03}$ and $D_{15}$ partial waves in Fig. 2, appear as positive bumps 
in the speed plots. In the $P_{01}$ and $P_{13}$ partial waves, one can
see that the resonant peak in the speed plot is broadened as compared
to the time delay peak which is narrower due to the presence of negative
time delay.  
That time delay can be negative even in the case of $\eta=1$ was noticed 
already by Wigner \cite{wigner} and an explicit example of this is given in 
\cite{lipschutz}. If $\eta \neq 1$, the situation described above still 
persists, but then, even the proportionality, 
$SP(E) = \vert \Delta t \vert /2$ vanishes. 
Indeed, we could have done our analysis without using the 
speed plots. However, since two earlier references used it in analyzing
the $KN$ data, we would like to make a correct comparison with their results.
\setcounter{equation}{0}
\section{Time delay, speed plots and resonances in $K^+N$ scattering}
\begin{figure}
\centerline{\vbox{
\psfig{file=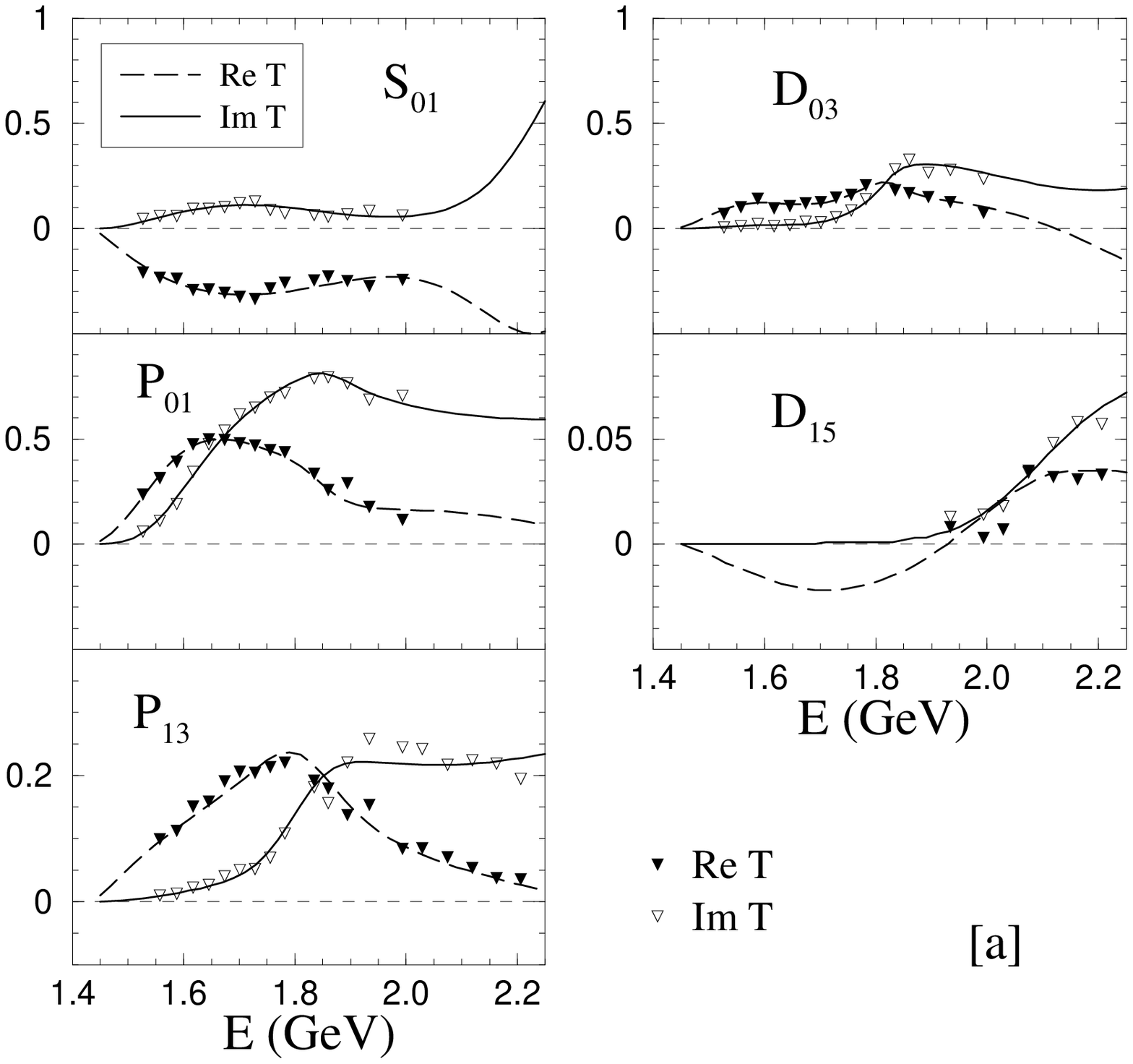,height=8cm,width=8cm}}}
\vskip0.1cm
\centerline{\vbox{
\psfig{file=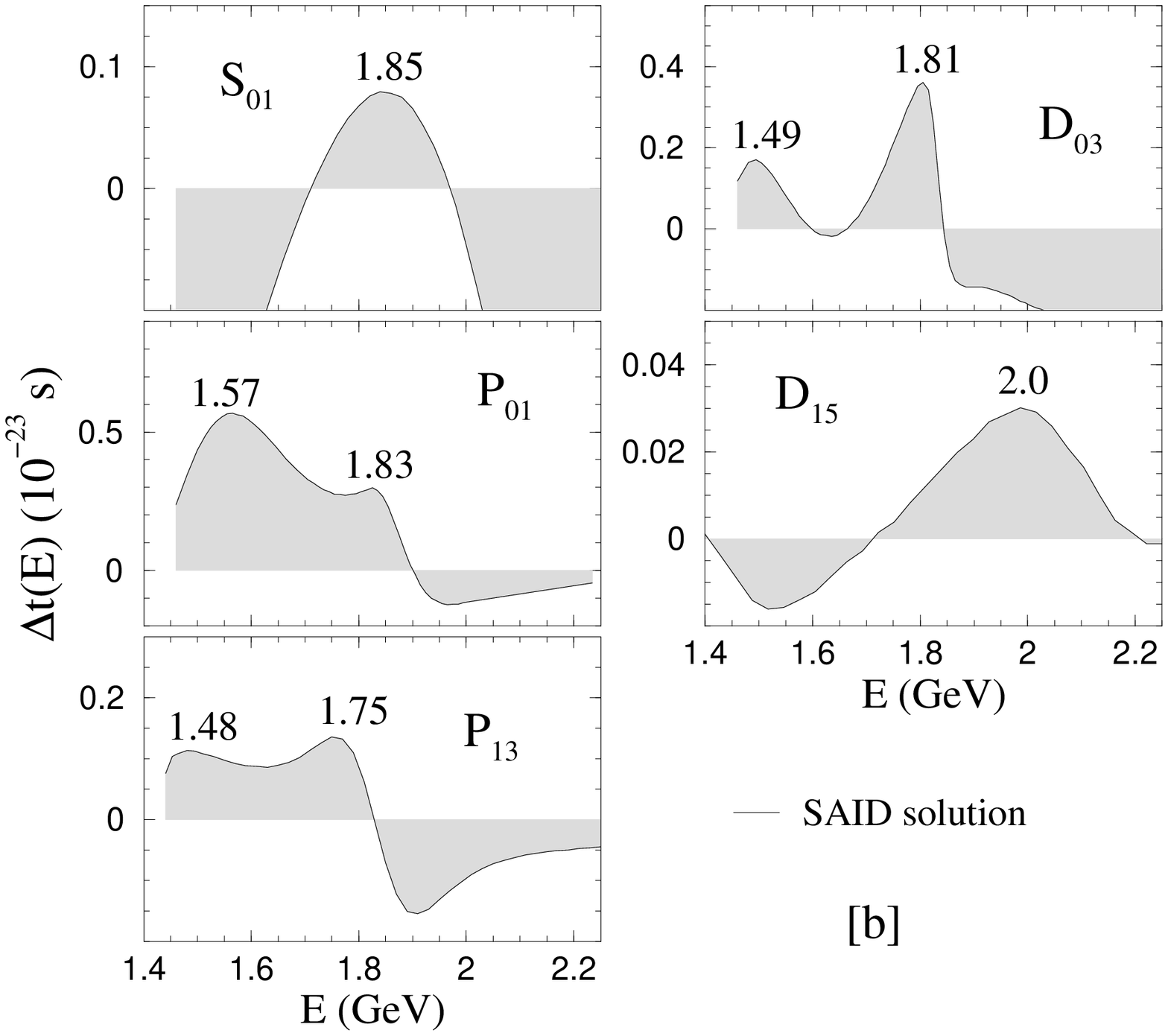,height=8.5cm,width=8cm}}}
\caption{(a) Real (dashed lines) and imaginary (solid lines) parts of the
$T$-matrix solutions \protect \cite{arndt} as compared to the single
energy values the real part (filled triangles) and imaginary part
(open triangles) for various partial waves in $K^+ N$ scattering
(b) time delay evaluated
using the $T$-matrix solutions (shown in (a))}
\end{figure}
Since the $T$-matrix (see Fig. 1) 
solutions found in \cite{arndt}, agree very well in the crucial cases
with the single energy values of the T-matrix and phase shifts, we calculate, 
with the exception of the $S_{01}$ partial wave 
(where we also use the single energy values), the time delay and 
speed plots from these solutions. The results for time delay are presented in 
Fig. 1 and for the speed plots in Fig. 2, where for comparison we 
have included the time delay results too. 
\begin{figure}
\centerline{\vbox{
\psfig{file=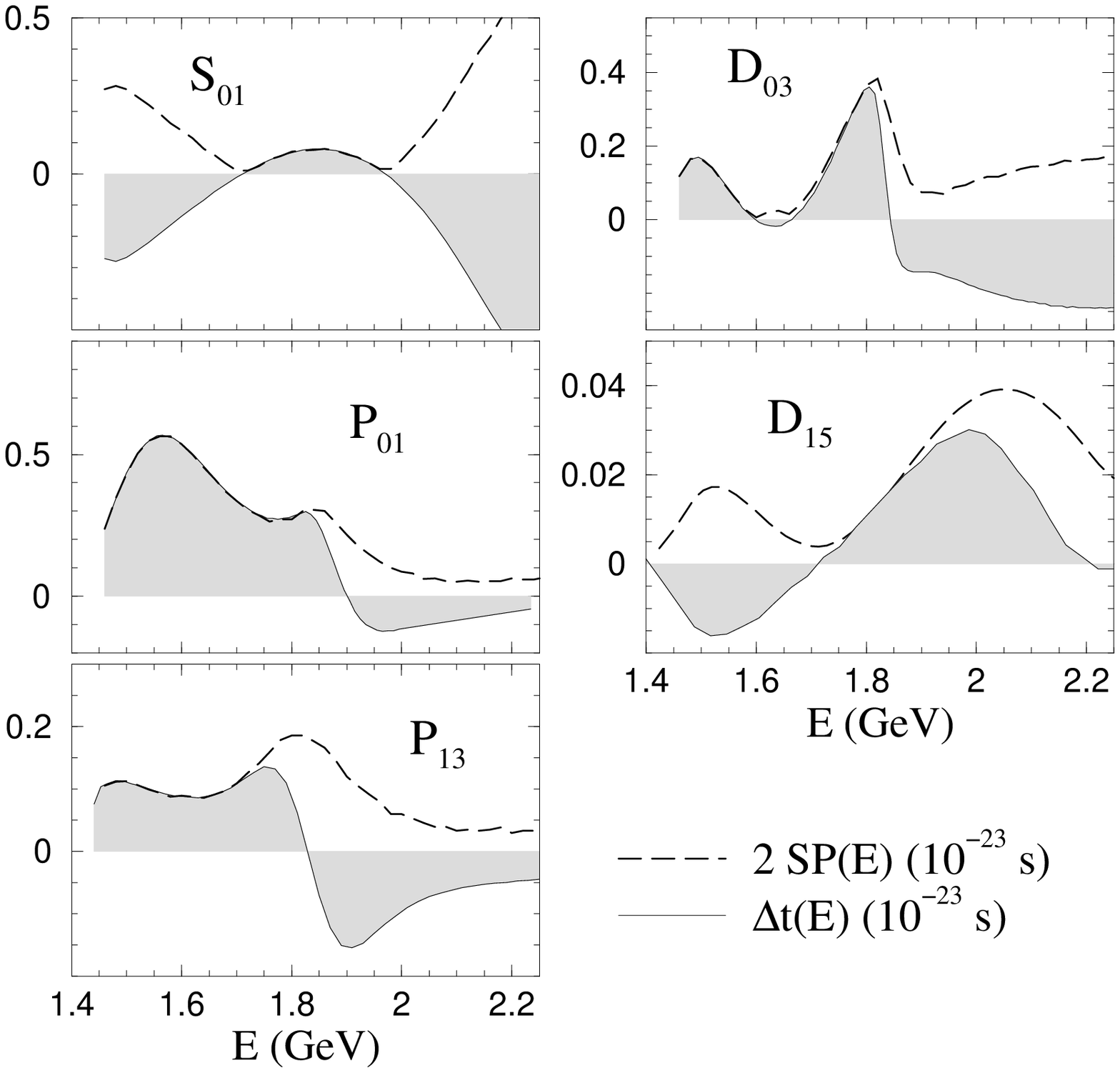,height=9cm,width=8cm}}}
\caption{Comparison of time delay and speed plots in various partial
waves of $K^+ N$ scattering.}
\end{figure}
It appeared to us that in the case of the partial wave,
$S_{01}$, the agreement between the solution and single energy values is 
not as good as in the other partial waves. We therefore decided to
calculate $\Delta t$ from the solution as well as from a fit made to the data 
points. These results can be found in Fig. 3.
\begin{figure}[h]
\centerline{\vbox{
\psfig{file=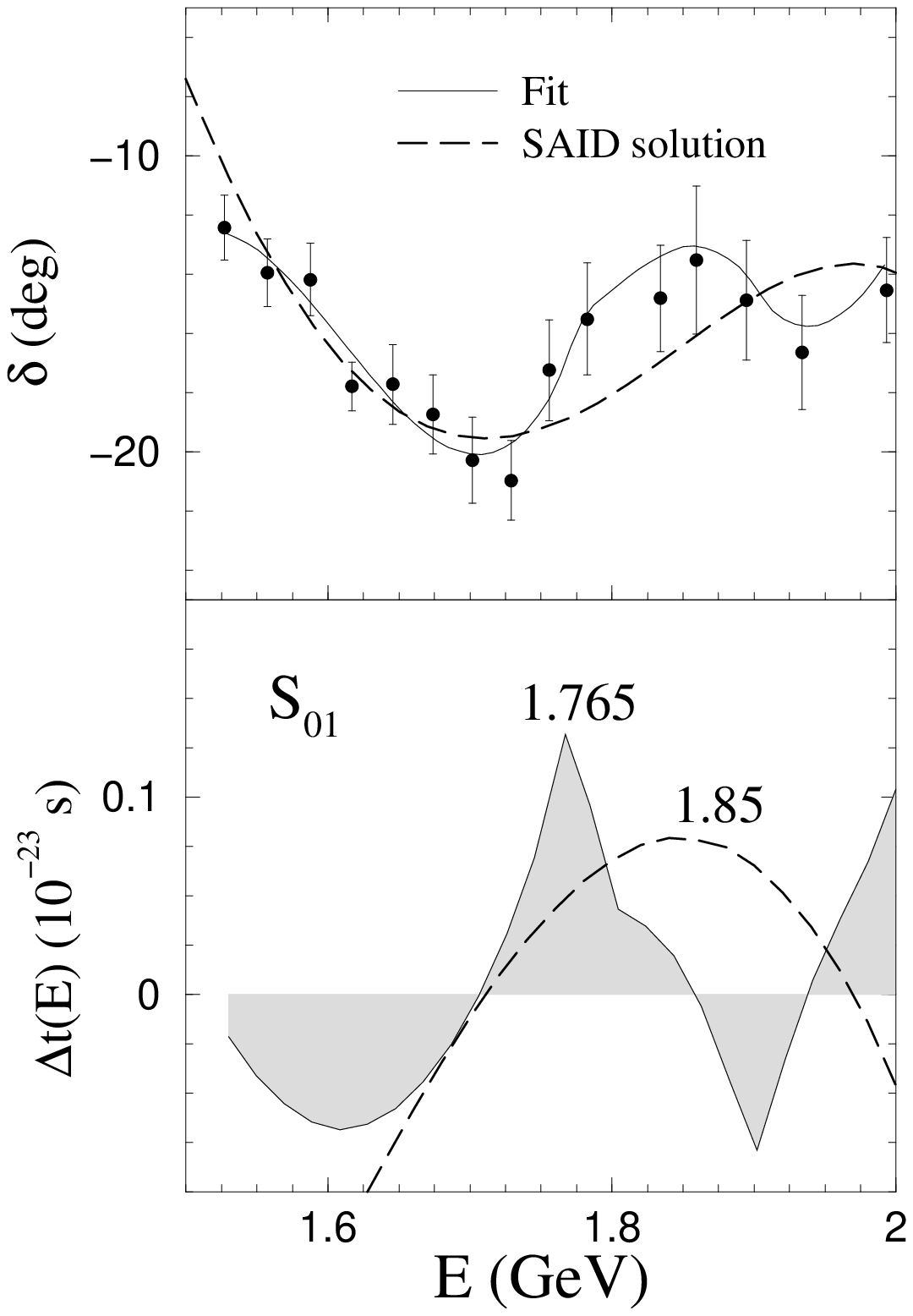,height=10cm,width=8cm}}}
\caption{Single energy values of phase shifts (filled circles) and
time delay evaluated from a fit (solid line) to these phase shifts
and from the solution (dashed line) as in Fig. 1, for the
$S_{01}$ partial wave in $K^+ N$ scattering.}
\end{figure}

The T-matrix poles found in \cite{arndt} are: $P_{01}(1831)$, $P_{13}(1811)$, 
$D_{03}(1788)$ and $D_{15}(2074)$. Comparing these values with the peak 
positions in the time delay plots for the corresponding partial waves, we see
that the agreement is very good. We do not expect a complete agreement here
between pole values and time delay peak values as this also does not occur
in the conventional cases of pion-nucleon resonances \cite{ng, ngandme}. As
evident from the theoretical example in \cite{lipschutz}, the mass parameters
appear shifted in time delay plots due to overlapping effects of  several 
resonances. The agreement of the peak values in the speed plot with the above 
mentioned pole values is also excellent. Hence we can ask ourselves the 
question: `what other objection prevents us from accepting the four 
$Z^*$ exotic resonances'? Note that the pole values (and of course 
the corresponding peaks in our plots) for $P_{01}$, $P_{13}$ and 
$D_{03}$ display a certain regularity. The three pole values are very 
close to the $K^*(892)N$ threshold. 
This is, however, a well known phenomenon. For instance, there are several well 
established pion-nucleon resonances
close to the $\rho N$ threshold. Theoretically, this phenomenon has been known
since the early 60's \cite{ball}.
In \cite{narod} (mentioned also in \cite{arndt2}) 
it is speculated that the signals for the resonances could be faked
by a $K^*$-box diagram (essentially the pion exchange diagrams
$K N \to K^*N$ and $K^*N \to KN$ glued together). 
In principle, 
we could address a similar question for higher lying pion-nucleon resonances
by replacing the $K^*$-box with a $\rho$-box, i.e. replacing the $K$ by $\pi$ 
and $K^*$ by $\rho$. 
  
Furthermore, in \cite{aaron} the very same
$K^*$-box was used as an {\it input} to dynamically generate the $Z^*$ 
resonances. The predicted $Z^*$ resonances in \cite{aaron} are $S_{01}$ and
$D_{03}$ {\it around} $1830$ MeV.
The model in \cite{aaron} has been put to a partial test in \cite{yogi}
which analyzes data by using this model. The speed plots in \cite{yogi} 
agree very well with our results. 
This actually means that, had the authors of \cite{yogi} evaluated time delay, 
their results would have also agreed with ours. 
It is important to note that our results for the time delay indicate that
the signal for a particular $Z^*$ resonance is a genuine one. To understand
this, we have to go back to the pion-nucleon resonances \cite{ng}.
In \cite{ng} it was found that the opening of a new channel
drives the positive time delay in the elastic channel into regions
of negative time delay. It actually starts becoming 
negative around the threshold of a new channel, very often interfering 
with the positive signal of the resonance itself. It was found that in
more than one case, this leaves only a small positive peak (due to the
resonance). That this is bound to happen is 
also clear from the connection of time delay with density of states (see
the discussion in \cite{ng} and \cite{ngandmeagain}). A `removal' of the 
initial states due to inelasticity makes the time delay negative. The
probability of this inelastic channel usually decreases with energy. 
Hence, in the {\it absence} of a resonance, for the $K^*$-box diagram, 
we would expect only a negative time delay 
around $K^*(892)N$ threshold becoming a positive {\it continuum} 
(due to the off-shell box diagram) at much higher energies.
We do not find such a behaviour in our time delay analysis, but
rather positive peaks around $1830$ MeV, followed by regions of
negative time delay. This is identical to the case of standard resonances in 
pion-nucleon scattering. Note that this
conclusion would be impossible to make using solely speed plots.

As mentioned in the Introduction, the rigor that one wishes to apply to 
exotics, demands that pole values be equal to the peak values of 
time delay and speed plots; a consistency not always encountered in 
the standard hadron resonances.
Hence, we shall only briefly discuss the other peaks found in our plots.
The peak in $S_{01}$ at $1850$ MeV (or $1765$ MeV in Fig. 3) was interpreted 
in \cite{aaron, yogi} as a resonance. In \cite{roiesnol} a 
resonance at $(1710)$ MeV in the $S_{01}$ partial wave was predicted. 
The time delay peaks at $1.5$ GeV, in the 
$P_{01}$, $P_{13}$ and $D_{03}$ partial waves, also have a certain
regularity. Some recent calculations \cite{weigel} made within the 
collective quantization scheme for chiral solitons predict a penta-quark
state, essentially an exotic $Z^*$ around mass 1570 MeV.  
In yet another chiral soliton model \cite{diakonov}, a $Z^*$ 
at $1530$ MeV was predicted. In ref. \cite{polyakov}, the authors show
that the $p p \rightarrow n \Sigma^+ K^+$ reaction provides optimal conditions
to detect the $Z^*$ if its mass is located around 1.5 GeV. Finally, we also
note that in a much older analysis \cite{nakajima}, 
one can see a clear peak around 1550 MeV in the speed plots of Fig. 3, 
although the authors
do not explicitly mention this fact in their table of resonance parameters. 
These predictions taken along with our findings of the low energy peaks
in time delay could possibly hint towards the existence of a low mass $Z^*$ 
in addition to the one around 1850 MeV. 
 
\setcounter{equation}{0}
\section{Conclusions}
The last entry of the $Z^*$ exotic resonances into the 
Particle Data Group compilation was in 
1992 \cite{1992}. In the same edition, it was remarked that it might take 
twenty years before the issue of the existence of the $Z^*$ resonances is 
settled. 
One might think that this is due to the lack of data. 
However, this is not the case. The first data sets, date back from 1969/1970,
followed by several others in the 70's and 80's with the last one in 1982.
The latest analysis is not restricted to a single data set, but includes
many of them \cite{arndt}. Hence we cannot blame the lack of data
if we are reluctant to decide the fate of the $Z^*$'s. 

In \cite{nakajima} it was remarked that: ``It was found that both data
showed reasonable agreement with each other but the agreement with the 
published analyses were not satisfactory.'' Reference \cite{nakajima}
is the same which does not find an agreement between pole value and 
values found
in speed plots (similar to \cite{martin}). Hence an internal consistency
of data and analysis is required. As far as the existence of resonances
is concerned, this can be done e.g. by checking if the pole values agree
with values obtained from time delay method and speed plots. We emphasize here
the time delay method since in contrast to speed plots the resonance region
in the time delay plots has to be positive 
\footnote{In principle, it could have happened that we find a peak in the
speed plots very close to the pole value. However, if in the time delay plot, 
it turns out that this peak is negative, we cannot attribute it
to a resonance.}. As remarked in the Introduction, this is more than we 
have for some of the four-star pion-nucleon resonances. 
Our analysis reveals that the pole values found in \cite{arndt} are in 
excellent agreement with the values for the masses found in time delay and
speed plots.

From a collection of data set progressing in time one would expect 
certain improvements, here  with regard to resonance extraction. 
In one of the early analysis of $KN$ scattering, it was demonstrated that 
`polarization measurements would be very helpful in demonstrating the
existence of these resonances'. At that time, only one
polarization measurement existed. However, 
this suggestion was picked up later in a series of experiments (see
\cite{arndt} for a list of references)
and included in the analysis in \cite{arndt}.

Finally, from different data and different analyses, one would demand a
certain consistency among each other \footnote{One should not overstress this
point; e.g. in the determination of the mass parameters, the latter are also 
not well determined for pion-nucleon resonances and different groups differ 
by as much as 10-20 \%}. Here comes a small surprise. Reference \cite{yogi}
analyzed the early data from 1970's using at the same time a
theoretical input to distinguish between solutions. This input is the
model in reference \cite{aaron} discussed in the previous section. Fig. 3 in
\cite{yogi} displays the speed plots for $S_{01}$, $P_{01}$ and $D_{03}$
partial waves in $KN$ collisions
calculated in 1973. Their form resembles very much our 
results for speed plots depicted in Fig. 2. More importantly, the peak values
of the speed plots in \cite{yogi} are in excellent agreement with ours,
though our result is based on a bigger data set and as far as we 
can say with lesser theoretical input. Yet another surprise comes, 
when we realize that the speed plot peaks in the 
$P_{13}$ and $D_{03}$ partial waves in \cite{nakajima} are around 
$1825$ MeV.  

To summarize, we can say that, we laid certain restrictions
for the existence of the exotic $Z^*$ resonances.
Within these restrictions the results were found to be
consistent, hinting towards the existence of these resonances.
We think that these hints should be taken seriously. 

\vskip0.5cm                                               
{\elevenbf \noindent Acknowledgements} \newline
We thank Juan Carlos Sanabria for stimulating discussions on exotic
baryons.
\\
\vskip0.5cm                                               
{\bf \it Note added after publication:} After this work was published, we
became aware of the work in \cite{nakano,clas,diana}, 
where excellent experimental evidence for an exotic $S=+1$ baryon,   
with mass around 1.54 GeV was reported. We speculate that it can
be identified with the $P_{01}(1.57)$ of the present work.

\end{document}